# Generalized electromagnetic energy-momentum tensor and scalar curvature of space at the location of charged particle

A.L. Kholmetskii[1], O.V. Missevitch[2] and T. Yarman[3]

[1]Belarus State University, 4 Nezavisimosti Avenue, 220030 Minsk, Belarus
[2]Institute for Nuclear Problems, Belarus State University, 11 Bobruiskaya Street, 220030 Minsk, Belarus
[3]Okan University, Akfirat, Istanbul, Turkey and Savronik, Eskisehir, Turkey

E-mail: khol123@yahoo.com

**Abstract** We consider the Einstein equation, where the common electromagnetic energy momentum tensor is replaced by its generalized equivalent as suggested in our earlier paper (A.L. Kholmetskii et al. Phys. Scr. 83, 055406 (2011)). Now we show that with this new electromagnetic energy-momentum tensor, the scalar curvature at the location of charges is significantly altered in comparison with the common result, and it even may change its sign. Some implications of the obtained results are discussed.

## 1. Introduction

In the present contribution we address to the Einstein equation and refer to our recent finding [1] that the common electromagnetic energy-momentum (EMEM) tensor is incomplete and is relevant only for the description of electromagnetic (EM) fields in the regions free of charged particles. For complete systems "charged particles along with their EM fields", an additional term should be introduced in the structure of EMEM tensor, which essentially influences its general properties. In particular, the trace of the new (generalized) EMEM tensor is no longer vanishing and thus we inevitably must conclude that the EM field should influence the scalar curvature of space unlike the common result. A closer look at this problem shows that in the spatial regions free of charged particles the scalar curvature of space is determined by the matter tensor as before. However, at the location of charges the contribution of the generalized EMEM tensor substantially influences the scalar curvature and may even change its sign in comparison with the common value (section 2). The result obtained is discussed in section 3.

## 2. Einstein equation with the modified electromagnetic energy-momentum tensor

We start with the Einstein equation

$$R_{\mu\nu} - \frac{1}{2}g_{\mu\nu}R = \frac{8\pi k}{c^4}T_{\mu\nu} \quad (\mu, \nu = 0\ldots3) \qquad (1)$$

where $R_{\mu\nu}$ is the Ricci tensor, **g** is the metric tensor, $R = g^{\mu\nu}R_{\mu\nu}$ is the scalar curvature of space, and $T_{\mu\nu}$ is the energy-momentum tensor, which represents the sum of the matter part and EM field part.

For a system of $N$ point-like charged particles the matter tensor has the form

$$T_{matter}^{\mu\nu} = \sum_{i=1}^{N} m_i \frac{dx_i^{\mu}}{dt} \frac{dx_i^{\nu}}{d\tau}, \quad (2)$$

(where $m_i$ being the mass density of particle $i$, and $\tau$ is the proper time), while the EMEM tensor is commonly written as

$$T_{EM}^{\mu\nu} = \frac{1}{4\pi}\left(-F^{\mu\gamma}F^{\nu}{}_{\gamma} + \frac{1}{4}g^{\mu\nu}F_{\gamma\alpha}F^{\gamma\alpha}\right), \quad (3)$$

where $F^{\mu\nu} = \partial^{\mu}A^{\nu} - \partial^{\nu}A^{\mu}$ is the tensor of EM field, and $A^{\mu}$ is the four-potential.

The trace of the EMEM tensor (3) is equal to zero and hence we arrive at the common result that EM field does not affect the scalar curvature [2], i.e.

$$R_{\mu}^{\mu} = R = -\frac{8\pi k}{c^4}\left((T_{matter})_{\mu}^{\mu} + (T_{EM})_{\mu}^{\mu}\right) = -\frac{8\pi k}{c^4}(T_{matter})_{\mu}^{\mu}. \quad (4)$$

However, in our recent paper [1] we have shown that the expression for the EMEM tensor (3) is incomplete, and it is applicable only to the spatial regions free of charged particles. In the general case, when the system in question includes both particles and their EM fields, eq. (3) has to be generalized to the form

$$T_{EM}^{\mu\nu} = \frac{1}{4\pi}\left(-F^{\mu\gamma}F^{\nu}{}_{\gamma} + \frac{1}{4}g^{\mu\nu}F_{\gamma\alpha}F^{\gamma\alpha}\right) - \frac{1}{c}(A^{\mu}j^{\nu})_{pr}, \quad (5)$$

where the subscript "pr" (proper) signifies that the four-potential components $A^{\mu}$ and four-current components $j^{\nu}$ are taken for the same source particle (particles).

In ref. [1], the incompleteness of eq. (3) has been demonstrated in two different ways:
- via the analysis of Poynting theorem for a bound (velocity-dependent) EM field;
- via the gauge modification of the canonical EMEM tensor to the symmetric form, assuming the presence of charges in the spatial region considered.

Concerning these ways, we refer the readers to the original work [1] for more details. Now we only mention that in the analysis of Poynting theorem we found its equivalent presentation for the case of purely bound EM field, which, however, cannot be reproduced with the common EMEM tensor (3) and with its any gauge modification. In contrast, the tensor (5) via its appropriate gauge transformation (named in ref. [1] as "gauge renormalization") yields both forms of the continuity equations mentioned above.

On the second way (the gauge transformation of the canonical EMEM tensor to the symmetric form) the incompleteness of EMEM tensor (3) follows from the common application of the homogeneous Maxwell equations in such a transformation (e.g. [2]), which, however are relevant only for the spatial regions free of charges. Using instead the general non-homogeneous Maxwell equations, we finally arrived to the EMEM tensor (5).

One can add that the equation (5) can be approached alternatively starting from the definition of the energy-momentum tensor given by Hilbert [1].

Further, we notice that the EMEM tensor (5) contains the four-potential in its last (additional) term and thus, it seems to be gauge-dependent. In this respect we have to mention that the four-divergence of time-like components of this term is equal to zero [1], i.e.
$\partial_{\mu}(A^{\mu}j^0) = 0$.

Therefore, the tensor (5) yields the standard Poynting theorem, like the common EMEM tensor (3).

Furthermore, as shown in ref. [1], via the operation of "gauge renormalization", the infinite (for point-like charges) term $-\frac{1}{c}(A^{\mu}j^{\nu})_{pr}$ is merged to the total observed mass of charged particle $M_t$, so that the motional equation derived with the tensor (5) does not include the four-potential and finally is gauge-independent. We point out that the possibility to include the term $-\frac{1}{c}(A^{\mu}j^{\nu})_{pr}$ to the total mass of charged particle in the correct mathematical way (i.e. gauge transformation of EMEM tensor (5)) is also important for the analysis presented below.

So, as we have mentioned above, the total energy-momentum tensor $T^{\mu\nu}$ of any macroscopic system of charged particles has to be understood as a sum of matter part and field part, i.e.

$$T^{\mu\nu} = T_{matter}^{\mu\nu} + T_{EM}^{\mu\nu}, \qquad (6)$$

where for a system of point-like particles $T_{matter}^{\mu\nu}$ is defined by eq. (2).

Now we notice an important property of EMEM tensor (5): in contrast to the conventional EMEM tensor (3), its trace is not equal to zero due to the contribution $-\frac{1}{c}(A^\mu j^\nu)_{pr}$. Hence using eq. (5) we get the possibility to describe the EM mass contribution into the total mass of charged particle which is not viable using the standard tensor (3). The latter statement issues from the known fact that any mass tensor must have the non-vanishing trace [3].

The *EM mass tensor* can be introduced by analogy with the matter tensor (2) as

$$\left(T_{EM}^{\mu\nu}\right)_{mass} = m_{EM} \frac{dx^\mu}{dt} \frac{dx^\nu}{d\tau} \qquad (7)$$

(where $m_{EM}$ is EM mass density), to which is added the Poincaré stresses tensor

$$T_P^{\mu\nu} = m_P \frac{dx^\mu}{dt} \frac{dx^\nu}{d\tau}, \qquad (8)$$

where $m_P$ is the negative mass density associated with the energy of "Poincaré stresses" necessary for the stability of the classical electron [4]. Thus the mechanical (matter) tensor (2) is modified to the form

$$T_{matter}^{\mu\nu} = (m + m_P) \frac{dx^\mu}{dt} \frac{dx^\nu}{d\tau}, \qquad (9)$$

and the total (observable) mass density of charged particle becomes the sum of three components:

$$m_t = m + m_P + m_{EM}, \qquad (10)$$

The detailed analysis of each of the components of eq. (10) for the classical electron has been carried out in ref. [4]. In particular, it has been found in [4] that the ratio

$$|M_t| : |M| : |M_P| : |M_{EM}| = 2 : 3 : 2 : 1 \qquad (11)$$

for any power function, describing the charge distribution "inside" the classical electron (here the capital letters stands for the corresponding mass component). Of course, in further application of the classical estimation (11) one has to take into account its conditional character. It is only important that all the mass components of this equation have the same order of magnitude, where $M_t$, $M$, $M_{EM}$ are positive, while $M_P$ is negative.

Thus the non-vanishing trace of the generalized EMEM tensor (5) allows us to introduce the concept of EM mass of classical charged particle in the non-contradictory way and, what is important for the purpose of the present paper, the fact of non-vanished trace of this tensor should be accounted in the solution of Einstein equation (1).

In particular, with the EMEM tensor (5), the expression for scalar curvature of space becomes

$$R = -\frac{8\pi k}{c^4} \left( (T_{matter})^\mu_\mu - \frac{1}{c}(A^\mu j_\mu)_{pr} \right) \qquad (12)$$

instead of eq. (4). In order to analyze this equation, we explicitly calculate the term

$$(A^\mu j_\mu)_{pr} = (\rho\varphi)_{pr} - (\mathbf{j} \cdot \mathbf{A})_{pr},$$

where $\rho$ is the charge density and $\varphi$ is the electric potential. For a system of $N$ charged particle this equation reads:

$$(A^\mu j_\mu)_{pr} = \sum_{k=1}^{N} \rho_{(k)} \varphi_{(k)} - \sum_{k=1}^{N} \mathbf{j}_{(k)} \cdot \mathbf{A}_{(k)}. \qquad (13)$$

Such a special form of eq. (13), where all quantities are evaluated for each fixed charge, signifies that we do not have a right to go from discrete distribution of charges characterized by the charge densities $\rho_{(k)}$ ($k=1...N$) to any averaged continuous charge distribution. (Here we recall that this operation is commonly used in general relativity theory with the introduction of continuous averaged distribution of matter). Therefore, we conclude that in any spatial point free of charged particles, the trace of EMEM tensor is equal to zero due to the equalities $\rho_{(k)}=0$, $j_{(k)}=0$ for any $k$. Hence the addition of the term $-\frac{1}{c}(A^\mu j^\nu)_{pr}$ to the common EMEM tensor does not affect the scalar curvature in any free spatial point, where the common Einstein equation (1) remains in force.

Thus, in comparison with common general theory of relativity, the scalar curvature is modified only within a region defined by particle's charge radius due to the non-vanishing trace of the component (13) of EMEM tensor, and this result seems insignificant in cosmology. Nevertheless, it is worth to continue the analysis of eq. (13), implying its possible usefulness to the classical limits of quantum gravity theories.

In such an analysis we will deal with a single charged particle ($N=1$), e.g. with the classical electron. Further on it is convenient to express the quantities in eq. (13) referred to the moving electron via the proper charge density $\rho_0$ and proper electric potential $\varphi_0$ fixed in its rest frame, assuming at the moment any gauge, where the four-potential is vanishing at the infinity. Then we get

$$\rho = \gamma\rho_0, \quad \varphi = \gamma\varphi_0, \quad j = \gamma j_0 = \gamma\rho_0 v, \quad A = \frac{\gamma\varphi_0 v}{c^2}, \qquad (14)$$

where $v$ is the velocity of the electron, and $\gamma$ is its Lorentz factor. Hence, substituting eqs. (14) into eq. (13), we obtain

$$\frac{1}{c}(A^\mu j_\mu)_{pr} = \gamma^2\left(\rho_0\varphi_0 - \rho_0\varphi_0\frac{v^2}{c^2}\right) = \rho_0\varphi_0. \qquad (15)$$

We mention that the product $\rho_0\varphi_0$ is always positive regardless of the sign of charge (electron, positron), and for the reasonable models of the classical electrons (see e.g. [4]), it represents the finite quantity, whose particular value depends on the adopted distribution of charge inside the classical electron. It is also interesting to notice that the term (15) responsible for the contribution of generalized EMEM tensor (5) into the scalar curvature, does not depend on the Lorentz factor $\gamma$.

Next we calculate the trace of the matter tensor (9):

$$(T_{matter})^\mu_\mu = (m+m_P)\frac{dx^\mu}{dt}\frac{dx^\mu}{d\tau} = \gamma(m+m_P)c^2 - \gamma(m+m_P)v^2 = \frac{(m+m_P)c^2}{\gamma}, \qquad (16)$$

where we have taken into account that for a moving particle $d\tau=dt/\gamma$.

Substituting eqs. (15), (16) into eq. (12), we derive the expression for spatial curvature at the location of classical electron:

$$R = -\frac{8\pi k}{c^4}\left(\frac{(m+m_P)c^2}{\gamma} - \rho_0\varphi_0\right). \qquad (17)$$

Further on we involve the equation obtained in ref. [4] for the density of Poincaré stresses mass:

$$m_P = -\frac{\rho_0\varphi_0}{c^2}. \qquad (18)$$

In a rough approximation (which is anyway allowed in the classical approach to particle physics), where we assume the homogeneous distribution of $m$, $m_P$ and $m_{EM}$ inside the electron, the relationship (11) can be adopted for the corresponding mass densities, too, i.e.

$$|m_t|:|m|:|m_P|:|m_{EM}| = 2:3:2:1. \qquad (19)$$

Then eqs. (18) and (19) allow us to express the product $\rho_0 \varphi_0$ via the density of total observable mass of the classical electron as

$$\frac{\rho_0 \varphi_0}{c^2} = m_t, \qquad (20)$$

and also to get the relationships

$$m_P = -m_t, \quad m = 3m_t/2. \qquad (21)$$

Hence, combining (17), (18), (20) and (21), we derive the expression for spatial curvature at the location of classical electron in gauge independent form:

$$R = \frac{8\pi k}{c^2} m_t \left(1 - \frac{1}{2\gamma}\right). \qquad (22)$$

We mention that the numerical coefficient of eq. (22) is, of course, model-dependent, and can be changed via a proper variation of the ratios of the mass density components (19). However, this circumstance is not so important, because in any case the classical analysis is just the approximation.

Anyway, we point out that the Lorentz factor $\gamma$ always exceeds unity, and thus the scalar curvature at the location of the classical electron is always positive. Hence it decreases the visible "size" of the electron in comparison with the case $R=0$.

We notice that the scalar curvature at the location of electron, being calculated with the common expression for the symmetric EMEM tensor (3) (whose trace is equal to zero), is completely determined by the matter tensor contribution and equal to

$$R = -\frac{8\pi k}{c^2} \frac{m}{\gamma},$$

which is always negative and does increase the visible "size" of electron.

In the present contribution we skip the classical analysis of hadrons, which occurs even more conditional that that for the classical electron. Anyway, the contribution of the last term in rhs of eq. (5) to the scalar curvature should be taken into account for the hadrons, too. In particular, we cannot exclude that the sign of the scalar curvature might be positive, too, just like for the electron.

## 3. Conclusion

Having modified the EMEM tensor (3) to the form (5), which covers the general case of the presence of both particles and fields, we substituted this new (generalized) EMEM tensor (5) into the Einstein equation (1). In the spatial regions free of charged particles, the trace of generalized EMEM tensor is equal to zero, just like with the common EMEM tensor. Hence we concluded that EM field does not affect the scalar curvature of space outside the charged particles. Moreover, since outside charged particles the term $-\frac{1}{c}(A^\mu j^\nu)_{pr}$ is equal to zero, its addition to the common EMEM tensor (3) does not affect the solutions of Einstein equation in such spatial regions. Thus the generalized EMEM tensor (5) we introduced in ref. [1], does not change anything in general relativity theory implications.

However, at the location of an electric charge the additional term $-\frac{1}{c}(A^\mu j^\nu)_{pr}$ becomes significant, and alters the scalar curvature inside the charged particles. In particular, for the classical electron (or positron) the scalar curvature not only changes its value, but the sign, too.

We hope that the results obtained in this paper could be useful for a test of classical limits of quantum theories of gravitation and, perhaps, for investigation of earliest stages of the Universe expansion.